\documentclass[prl,twocolumn,preprintnumbers,amsmath,amssymb]{revtex4}

\usepackage{epsfig}
\usepackage[colorlinks=true,linkcolor=blue]{hyperref} 
\usepackage{color}
\usepackage{ulem}
\usepackage{dcolumn}
\DeclareGraphicsExtensions{.pdf}

\begin{document}

\title{Signature of a light charged Higgs boson from top quark pairs at the LHC}

\author{YaLu~Hu$^{1,2}$,~ChunHao~Fu$^{1,2}$,~Jun~Gao$^{1,2}$}

\affiliation{
    $^1$ INPAC, Shanghai Key Laboratory for Particle Physics and Cosmology \\ \& School of Physics and Astronomy, Shanghai Jiao Tong University, Shanghai 200240, China \\
    $^2$Key Laboratory for Particle Astrophysics and Cosmology (MOE), Shanghai 200240, China
    }

\begin{abstract}
The charged Higgs boson is a smoking gun of extensions
of the standard model with multiple Higgs-doublets, and has been searched for at various collider
experiments.
In this paper, we study signature of a light charged Higgs boson produced by top quark pairs
at the LHC, with subsequent three-body decays into a $W$ boson and a pair of bottom quarks.
Cross sections on final states of two $W$ bosons plus four bottom 
quarks have been measured by the ATLAS collaboration at the LHC 13 TeV. 
We reinterpret the experimental data under the scenario of a light charged Higgs boson,
and find improved agreements. 
We obtain the first limit from LHC direct searches on the total branching ratio of the
three-body decay, $Br(t\rightarrow H^+b)\times Br(H^+\rightarrow W^+b\bar b)$,
and the strongest direct constraints on the parameter space
of a class of type-I two-Higgs-Doublet models. 

\end{abstract}
\maketitle

\begin{figure}[h!]
	\begin{center}
		\includegraphics[width=0.43\textwidth]{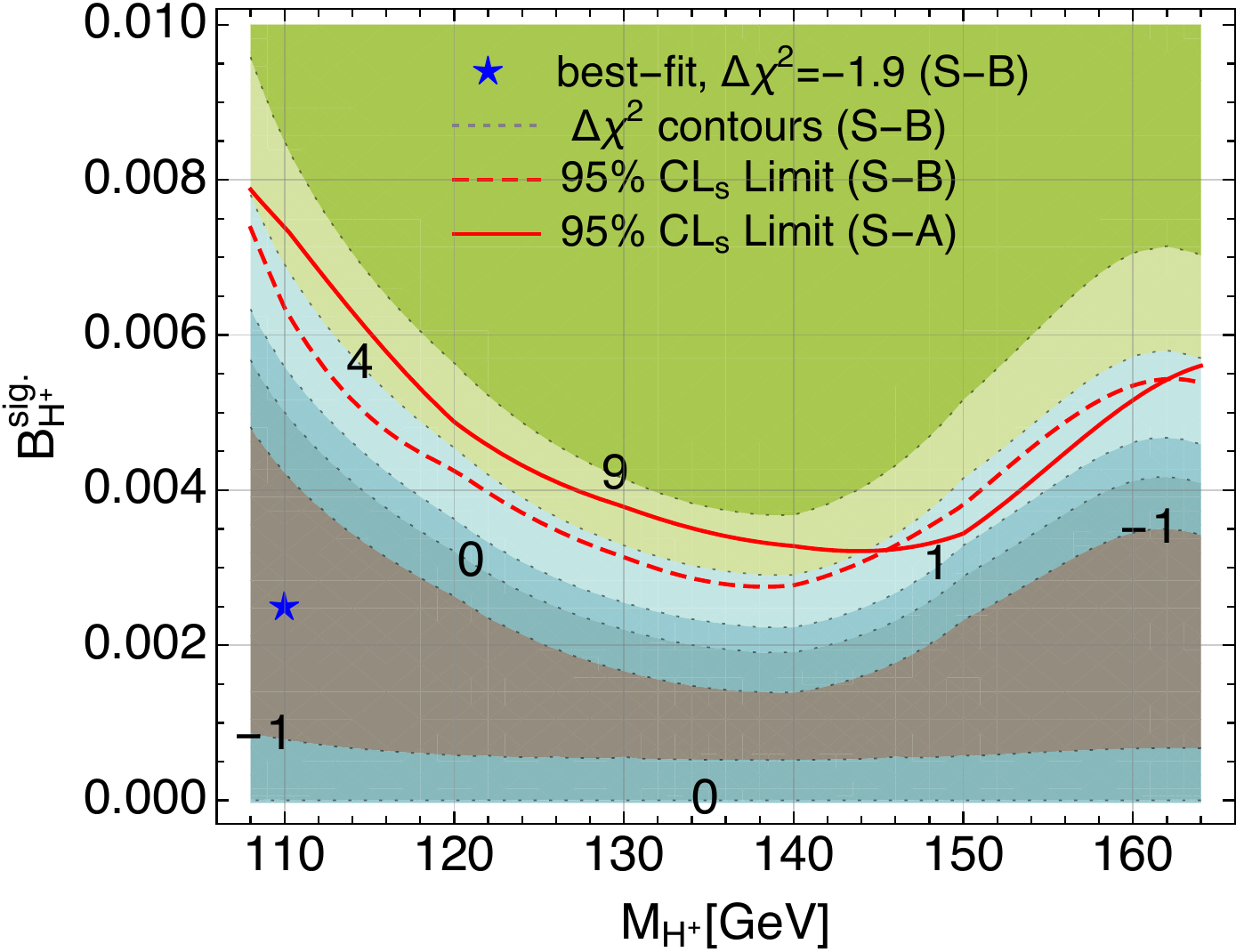}
	\end{center}
	\caption{\label{fig:chi2}
$\chi^2$ contours on the plane of the mass of the charged Higgs boson
and the signal strength for the scenario $\mathcal {B}$,
with $\chi^2$ of the pure SM predictions subtracted.
95\% CLs limits on the signal strength for the two scenarios of the
charged Higgs boson as functions of the mass are also shown.} 
\end{figure}

\pagebreak
\newpage
\noindent \textbf{Introduction.}
The successful operation of the CERN Large Hadron Collider (LHC) and
the ATLAS and CMS experiments has
led to the discovery of the Higgs boson, the final piece of the
standard model (SM)~\cite{Chatrchyan:2012ufa,Aad:2012tfa} of particle physics.
Ten years after the discovery it remains mysterious: how the masses of
fermions are arranged in a pattern of hierarchy, and if the SM Higgs boson
is the only elementary particle with spin zero.
A natural extension of the SM Higgs sector is to include multiple Higgs-doublets,
for instance, the well motivated
two-Higgs-Doublet models (2HDMs)~\cite{Lee:1973iz,Branco:2011iw}.
The charged Higgs boson is a smoking-gun signature of such new physics models
and has received a lot of attentions recently. 
Direct searches on the charged Higgs boson have been carried out previously
at LEP~\cite{1301.6065}, Tevatron~\cite{hep-ph/0510065,1104.5701}, and now
at the LHC~\cite{2102.10076,2001.07763}.
Constraints on branching ratios of
various decay channels of the charged Higgs boson have been obtained,
depending on its mass. 
We expect improved sensitivities at the upcoming run of the LHC with
high luminosities, which may lead to a discovery of the charged Higgs boson
or even stringent limits.
The searches of a light charged Higgs boson at the LHC, namely with mass smaller
than difference of the masses of the top quark and the bottom quark, benefit
from the large production cross sections of the top quark pair which decays into
the charged Higgs boson and a bottom quark.
A light charged Higgs boson can undertake several possible
decays~\cite{1810.09106,1912.10613,2012.09200,2202.03522}, for instance
in the 2HDMs into $\tau^+\nu_{\tau}$, $c\bar s$, $c\bar b$, $t^{(*)}\bar b$, and
$W^{+}A$ if the mass of the CP-odd Higgs boson $A$ is below the threshold.
The latter two decays can both lead to $W^+b\bar b$ final states either
directly or via cascade decay of $A$, which can be dominant over the
two-body decays.
There exist many theoretical studies demonstrating potential
of the $W^+b\bar b$ decay channel on exploring the parameter space of the
2HDMs~\cite{Akeroyd:1998dt,1607.02402,1906.02520,2103.07484,2106.13656,2201.06890,2206.02439}.
However, to our best knowledge, there are no dedicated experimental searches
reported so far at the LHC on this three-body decay channel.
In this Letter, we utilize a measurement on inclusive and differential fiducial
cross sections of final states with two $W$ bosons and four bottom quarks
by the ATLAS collaboration at the LHC 13 TeV with an integrated
luminosity of 36~$fb^{-1}$~\cite{1811.12113}.
We reinterpret the experimental data under the scenario of a light charged Higgs boson,
and find improved agreements with the data. 
Here one of the top quarks decays into a charged Higgs boson and the
other follows the SM decay into a $W$ boson and a bottom quark.
We obtain the first limit from LHC direct searches on the total branching fraction of the
three-body decay, $Br(t\rightarrow H^+b)\times Br(H^+\rightarrow W^+b\bar b)$, for a charged Higgs boson with
mass smaller than the top quark.
Our result sets strong constraints on the parameter space of type-I 2HDMs. 
\noindent \textbf{Theory and data comparison.}
ATLAS measured final states including a pair of $W$ bosons and four bottom quarks
and compared with SM predictions from QCD production of a pair of top quarks
associated with a pair of bottom quarks with subsequent decays.
The measurement has been separated into the pure leptonic channel where one of
the $W$ bosons decays into an electron and the other into a muon, and the lepton
plus jets channel where one of the $W$ bosons decays into jets and the other into
either a muon or electron~\cite{1811.12113}.
We refer to them as leptonic and jet channel for simplicity in the following sections.  
The $W$ boson can decay to the electron and muon either directly or via an
intermediate $\tau$-lepton, with the latter contributes about 10\%
to the cross sections.
For each channel the signal region is further classified as that with
at least four $b$-jets and that with at least three $b$-jets, since
one of the bottom quarks can be out of experimental acceptance.
The results on inclusive and differential fiducial cross sections
have been unfolded to particle level to ensure comparisons with
theoretical predictions from MC event generators.
Detailed definitions on the fiducial region can be found in the experimental
publication~\cite{1811.12113} and are also implemented in the public
Rivet~\cite{1003.0694} analysis routine.
We note there is another measurement on similar final states by the CMS
collaboration~\cite{1909.05306}.
However, we conclude that the CMS measurement can not be used in our
study of the charged Higgs boson since it requires reconstructions of
the top quarks following the SM decay mode.
Theoretical predictions on the binned cross sections in the presence
of a light charged Higgs boson can be expressed as
\begin{align}\label{eq:sig}
	\sigma_{pre}^{bin}&\equiv\sigma_{SM}^{bin}+\sigma_{H^+}^{bin}\\
	    &=\sigma_{SM}(t\bar t b\bar b)\epsilon^{bin}_{SM}
	    +2B^{sig}_{H^+}\sigma_{SM}(t\bar t)\epsilon^{bin}_{H^+} \nonumber
\end{align}
assuming the branching ratio of the top quark decays into the charged
Higgs boson is small.
$\sigma_{SM}(t\bar t b\bar b)$ is the SM cross section on QCD
production of four heavy quarks, and $\epsilon^{bin}_{SM}$ represents
the efficiency for prescribed kinematic bin including branching fractions
of SM decays of the $W$ boson.
There are also other SM processes contributing to the same final states
which have been subtracted from the experimental data already.
Contributions from the charged Higgs boson have been factorized into the
SM cross section on QCD production of top quark pair, the efficiency,
and the total branching fraction $B^{sig}_{H^+}$ for the
full decay chain of $t\rightarrow H^+ b \rightarrow W^+b b\bar b$.
The factor of 2 in Eq.~(\ref{eq:sig}) is due to the fact that the charged Higgs
boson arises from decays of both the top quark and anti-quark.
Non-resonant production cross section of a light charged Higgs boson
is generally below 10\% of the cross section for on-shell production of the
top quark pair with subsequent decays~\cite{1607.05291}.
We do not include the non-resonant contributions for simplicity. 
We treat $B^{sig}_{H^+}$ as an input of the signal strength of new physics, and
derive the efficiency $\epsilon^{bin}_{H^+}$ from MC simulations that only depends
on masses of the Higgs bosons.
We generate event samples with MG5\_aMC@NLO~\cite{1405.0301} followed by parton showering (PS) and
hadronizations with PYTHIA8~\cite{Sjostrand:2014zea} in the four flavor number scheme (4FS), and
analyse the events with the public routine of the ATLAS analysis in Rivet~\cite{1003.0694}.
We use CT18 PDFs~\cite{1912.10053} and a top (bottom) quark pole mass of
172.5 (4.75)~GeV in simulations, and set the default renormalization
and factorization scales to the sum of transverse energy of all final states divided by two.
For MC simulations of the charged Higgs boson, we use a model file
of the general 2HDMs generated with FeynRules~\cite{Alloul:2013bka}.
The efficiency $\epsilon^{bin}_{H^+}$ is calculated with event samples generated at leading
order in QCD matched with PS.
We set the total cross section of SM top-quark pair production to 838.5~pb at LHC 13 TeV,
calculated with Top++2.0~\cite{1303.6254,1112.5675} at next-to-next-to-leading order (NNLO) and next-to-next-to-leading
logarithmic accuracy in QCD.
We include two scenarios on the three-body decay of the charged Higgs boson with a mass
between 110 and 160~GeV.
In the first case, scenario $\mathcal{A}$, we assume that the additional neutral Higgs bosons
are heavier than $H^+$ and the three-body decay goes directly as $H^+\rightarrow t^{(*)}\bar b
\rightarrow W^+b\bar b$.
In scenario $\mathcal{B}$, we assume a CP-odd Higgs boson $A$ with a mass $M_A=M_{H^+}-85$~GeV,
similar to one in Ref.~\cite{1905.07453}.
The three-body decay goes through a cascade $H^+\rightarrow W^+A(b\bar b)$.
For the SM predictions, we calculate $\sigma_{SM}^{bin}$ directly using event
samples of $t\bar t b\bar b$ generated at next-to-leading order (NLO) in QCD
matched with parton showering.
We find good agreements of our SM predictions with those predictions shown in the ATLAS
analysis~\cite{1811.12113}.
In the remaining part of our study we instead use the theoretical predictions reported
in the ATLAS analysis directly since comprehensive estimations of theoretical uncertainties are available. 
Be specific, for the inclusive fiducial cross sections, we take the theoretical
predictions from SHERPA2.2~\cite{0811.4622} at NLO+PS in 4FS.
The theoretical uncertainties are
obtained by varying the renormalization and factorization scales by factors of 0.5 and 2.0
and including PDF uncertainties from NNPDF3.0 NNLO PDFs~\cite{1410.8849}.
For normalized fiducial distributions, we take the predictions from
POWHEG+PYTHIA8 in the 4FS for $t\bar t b \bar b$ production~\cite{1802.00426},
and in the 5 flavor number scheme (5FS) for $t\bar t$ production~\cite{0707.3088},
both at the accuracy of NLO+PS. 
In the 5FS, there are three predictions with different tunes of POWHEG and
PYTHIA8~\cite{ATLAS:2016xpx}, and additional bottom quarks are generated from PS.
We take midpoints of the envelope of the four predictions as the central
prediction and half width of the envelope as theoretical uncertainties.
The PDF uncertainties are negligible for normalized distributions and are
thus not included.
We have checked the choice of the nominal theoretical prediction on
the SM $t\bar t b\bar b$ production has little impact on our final results.
There are also several recent calculations on the SM
predictions at NLO+PS accuracy~\cite{1408.0266,1709.06915},
and at NLO including full non-resonant and off-shell contributions with leptonic
decays of the $W$ bosons~\cite{2008.00918,2105.08404,2202.11186}.
\begin{figure}[h!]
	\begin{center}
		\includegraphics[width=0.43\textwidth]{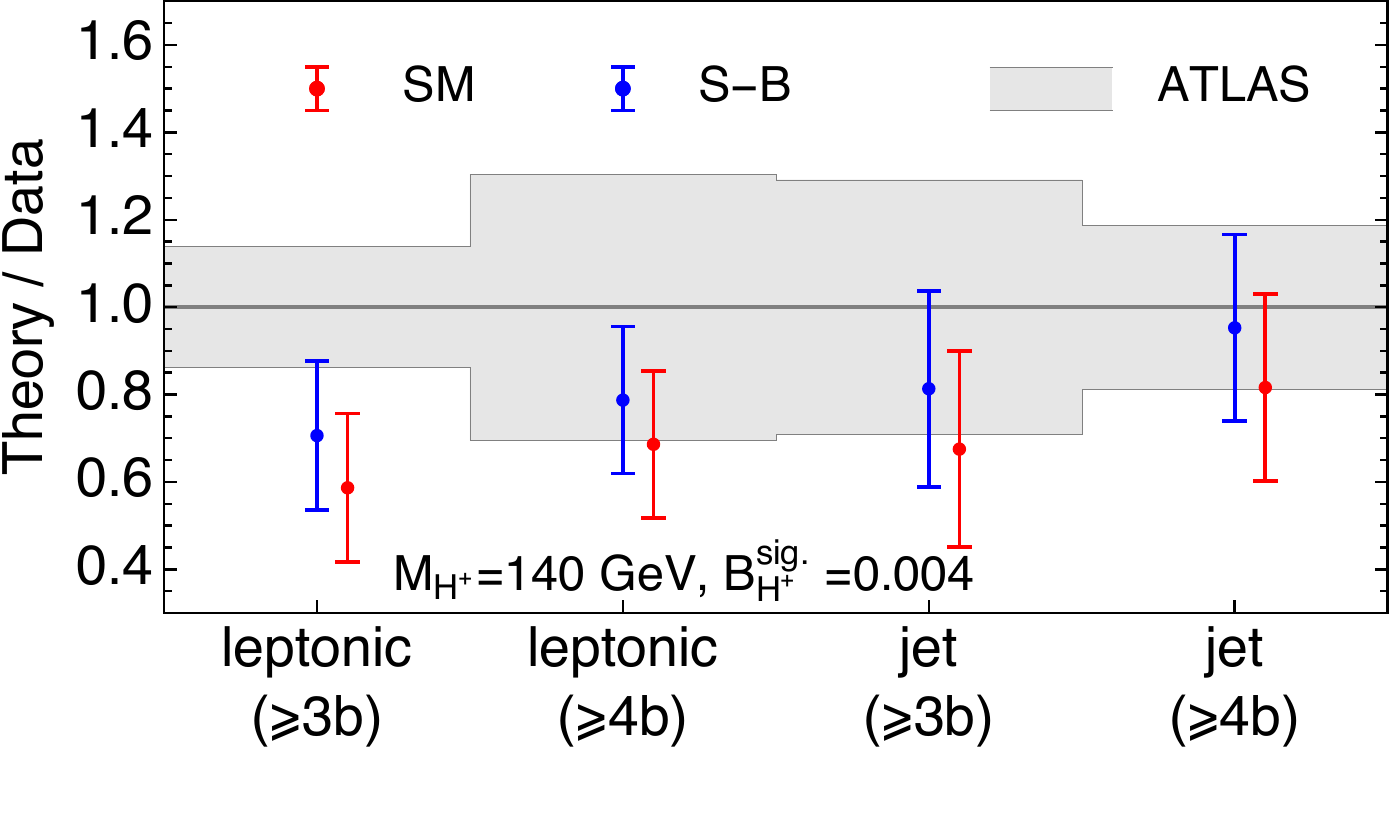}
		\includegraphics[width=0.43\textwidth]{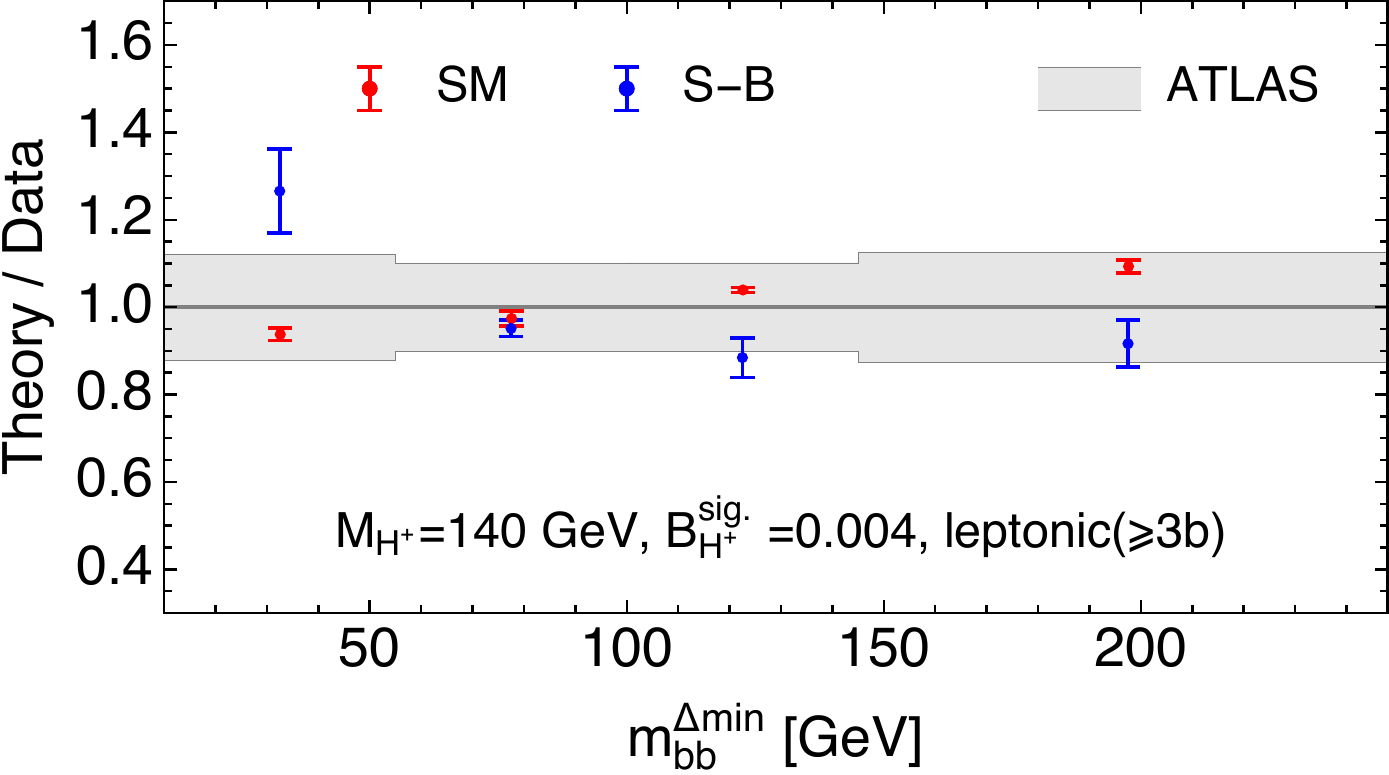}
	\end{center}
	\caption{\label{fig:sig}
Comparisons of theory and data on the inclusive fiducial cross
sections and the normalized $m^{\Delta{min}}_{bb}$ distribution of the leptonic
channel, for the SM and the SM with a charged Higgs boson
of scenario $\mathcal{B}$ in addition.
We take a mass of 140~GeV and a signal strength of 0.4\% for demonstration.
Theory predictions are normalized to the central measurements.
The bands and error bars represent the total experimental and theoretical
errors respectively.}
\end{figure}
We perform a survey on the inclusive and various differential fiducial
cross sections measured by ATLAS and select three data sets.
The first set is on the inclusive fiducial cross sections of the leptonic and jet channel
with three and four $b$-jets respectively. 
The other two are normalized distributions of invariant mass of the pair of
two closest $b$-jets in $\Delta R$, $m^{\Delta{min}}_{bb}$, for the leptonic
and jet channel respectively.
Distributions of $m^{\Delta{min}}_{bb}$ are most sensitive to the
charged Higgs boson which tends to generate a smaller invariant mass for the
two closest $b$-jets.
We drop the last bin in each of the two distributions since they are not
independent from the rest for normalized distributions. 
We show comparisons between theory and data on the inclusive fiducial cross
sections and the normalized $m^{\Delta{min}}_{bb}$ distribution of leptonic
channel in Fig.~\ref{fig:sig}, for the SM and the SM with a charged Higgs boson
of scenario $\mathcal{B}$ in addition.
We take a mass of 140~GeV of the charged Higgs boson and a signal
strength of 0.4\% for demonstration.
Both theoretical predictions are normalized to the central measurements.
The bands and error bars represent the total experimental and theoretical
errors respectively.
For the normalized distributions, the theoretical uncertainties increase
when including contributions from the charged Higgs boson, due to the
uncertainties propagated from the SM inclusive fiducial cross
sections.
We find the charged Higgs boson brings fiducial cross sections closer to central
of the ATLAS data and leads to a significant increase of $m^{\Delta{min}}_{bb}$ distribution
in the first bin.   
The log-likelihood function $\chi^2$ summed over the chosen data sets
is calculated as
\begin{align}
	\chi^2(M_{H^+}, B^{sig}_{H^+})=\sum_i\frac{(\sigma^{bin, i}_{pre}-\sigma^{bin, i}_{dat})^2}
	{\delta_{sta, i}^2+\delta_{sys, i}^2+\delta_{th, i}^2},
\end{align}
where the denominator includes statistical, systematic
and theoretical uncertainties, respectively.
$\sigma^{bin, i}_{dat}$ is
the central measurement of the cross section in the $i$-$th$ bin.
We can not include distributions on different kinematic variables at once
since they are fully correlated in statistics.
We plot $\chi^2$ contours on the plane of
the mass of the charged Higgs boson and the signal strength in Fig.~\ref{fig:chi2},
for the scenario $\mathcal {B}$ with a total number of
experimental data points of 12.
The results are similar for the scenario $\mathcal {A}$ which are not shown here.
We have subtracted the $\chi^2$ of pure SM predictions in the contours, which is
6.9 units.
For both scenarios the best-fit is found at $M_{H^+}$ close to
110~GeV and with a total branching fraction $B_{H^+}^{sig}$ of
about $0.3$\%$\sim 0.4$\%.
The $\chi^2$ is lower by about 2 units compared to the SM case.
Inclusion of contributions from the charged Higgs boson leads to moderate
improvement on description of the data especially because of enhancements
to the inclusive fiducial cross sections.
We can deduce upper limits on the total branching fraction of
$t\rightarrow H^+ b \rightarrow W^+b b\bar b$ for fixed values
of $M_{H^+}$.
We use the CLs method~\cite{AL2002Presentation} together with the 
$\chi^2$ function.
Note here we only include the two date sets of normalized distributions
(8 bins in total) for calculations of the $\chi^2$.
Upper limits on $B^{sig}_{H^+}$ for a fixed $M_{H^{+}}$ at a confidence
level 1-$\alpha^{\prime}$ is determined as
\begin{equation}
\hat{\mu}+\Delta_{\mu} \Phi^{-1}\left(1-\alpha^{\prime} \Phi(\hat{\mu}
/ \Delta_{\mu})\right),
\end{equation}
where $\hat{\mu}$ is the best-fit of $B^{sig}_{H^+}$ with fixed $M_{H^{+}}$,
and $\Delta_{\mu}$ is the uncertainty estimated by requiring $\Delta\chi^2=1$
compared to the best-fit.
$\Phi$ is the cumulative distribution function of normal distribution.
We plot the 95\% CLs limits for the two scenarios of the charged Higgs boson
as functions of the mass in Fig.~\ref{fig:chi2}. 
The best limit is about 0.32(0.28)\% at a mass of about 140~GeV
for the scenario $\mathcal{A(B)}$, and the limits deteriorate 
at both ends of the mass range.
The results represent first limits on the branching
fraction of such decay channel from direct searches at the LHC.
Furthermore, we emphasize that in deriving the limits we
have only used a single kinematic distribution in $m^{\Delta{min}}_{bb}$.
We expect enhanced sensitivity to the charged Higgs boson in future
experimental analyses once multivariate discrimination methods are used. 
\noindent \textbf{Model constraints.}
We can translate above limits into constraints on the parameter space of the
relevant 2HDMs.
We take type-I 2HDMs as examples which are less constrained by direct
searches at the LHC~\cite{1906.02520}.
One important feature of type-I models is that couplings of the charged
Higgs boson and the CP-odd neutral Higgs boson to fermions are both
proportional to masses of the fermions divided by $\tan\beta$.
The latter is
the ratio of vacuum expectation values of the two Higgs doublets.
The total branching ratio of the three-body decay can be written as
\begin{align}
B_{H^+}^{sig}(\mathcal{A})&=Br(t\rightarrow H^+b)\times Br(H^+\rightarrow t^{(*)}\bar b),\\
B_{H^+}^{sig}(\mathcal{B})&=Br(t\rightarrow H^+b)\times Br(H^+\rightarrow W^+A)\nonumber \\
	&\times Br(A\rightarrow b\bar b), \nonumber
\end{align}
for scenario $\mathcal{A}$ and $\mathcal{B}$, respectively.
In 2HDMs of type-I, the branching ratio $Br(t\rightarrow H^+b)$ only depends
on $M_{H^+}$ and $\tan \beta$.
For that we calculate the partial widths of the top quark at NNLO in QCD
for both its SM decay~\cite{1210.2808} and the decay into the charged
Higgs boson~\cite{2201.08139}.
For decays of the charged Higgs boson there are competing channels
including $\tau^+ \nu_{\tau}$, $c\bar s$, $t^{(*)}\bar b$ and $W^+A$, while others are
negligible in 2HDMs of type-I.
The relative weights of fermionic channels only depend on the mass of the charged
Higgs boson, where $\tau^+\nu_{\tau}$ ($t^{(*)}\bar b$) is dominant at small (large)
$M_{H^+}$. 
For simplicity, we assume the CP-odd Higgs boson $A$ is heavier than
the charged Higgs boson in models associated with scenario $\mathcal{A}$
to avoid the decay into $W^+A$. 
In models related to scenario $\mathcal{B}$ where $M_{H^+}-M_A$ is fixed to 85~GeV, 
the branching
fraction to $W^+A$ is almost 100\% for the parameter space of interests, namely with
$\tan\beta\gtrsim 1$.
Decays of the Higgs boson $A$ to fermions are dominant.
The branching fraction is 82(0.024)\% to a pair of
bottom quarks (muons) and is almost independent
of $\tan \beta$ and $M_{H^+}$ for the parameter space considered.
From above we conclude both $B_{H^+}^{sig}$
depend only on $M_{H^+}$ and $\tan\beta$ for which we will set constraints on.
We calculate the branching fractions of decays of $H^+$ and $A$ in type-I models using
the 2HDMC-1.8.0 program\cite{0902.0851}.
\begin{figure}[h!]
	\begin{center}
		\includegraphics[width=0.48\textwidth]{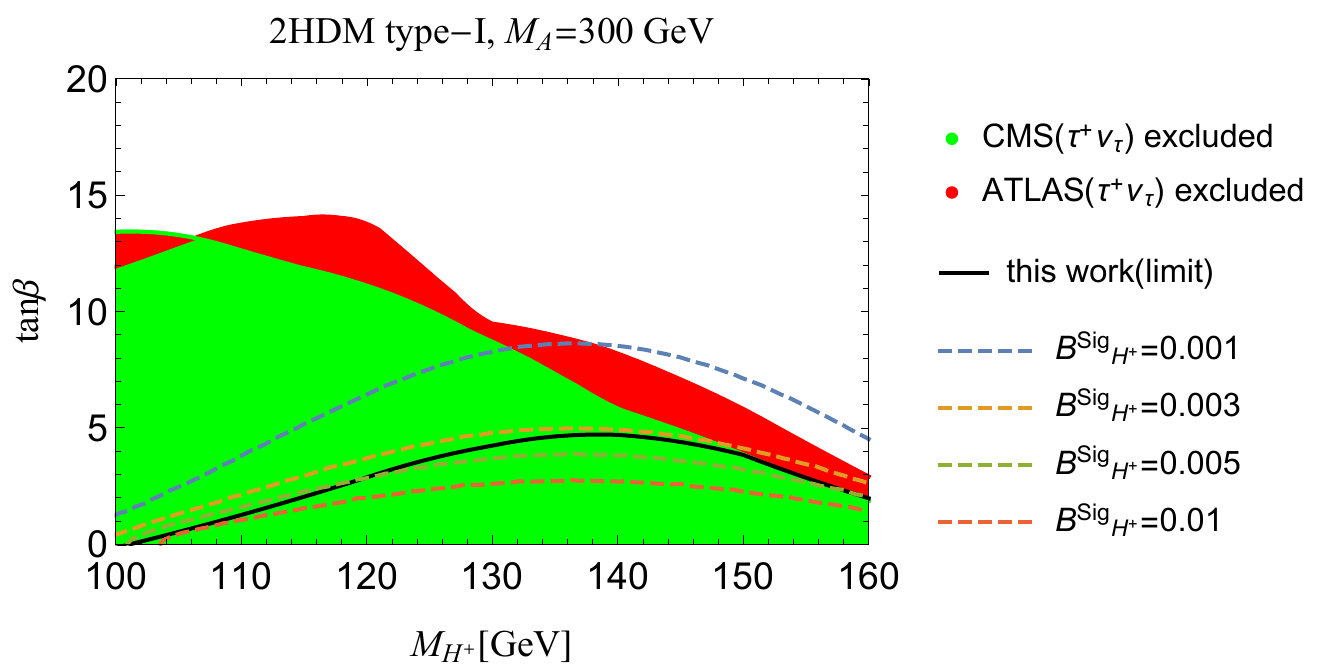}
		\includegraphics[width=0.48\textwidth]{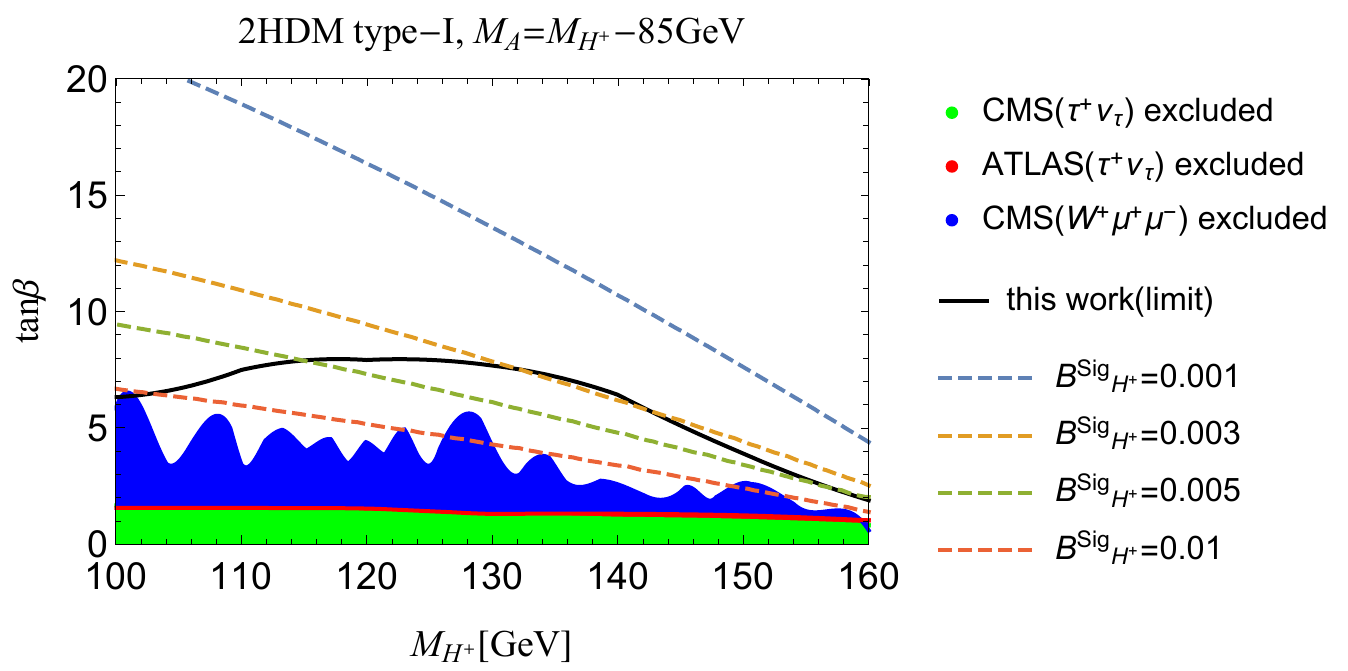}
	\end{center}
	\caption{\label{fig:para}
Excluded regions of the parameter space by various direct searches of the
charged Higgs boson at the LHC for the two classes of type-I models considered.
Dashed lines represent contours of the respective signal strength ranging from
0.1\% to 1\%.
Regions under the solid curves are excluded by this work.} 
\end{figure}
We reproduce constraints on the parameter space imposed by previous
direct searches of a light charged Higgs boson at the LHC for comparisons, all at 13 TeV with
an integrated luminosity of about 36~$fb^{-1}$.
They include searches for $\tau^+\nu_{\tau}$ decay channel
from both CMS~\cite{1903.04560} and ATLAS~\cite{1807.07915}.
In addition, there exists a search for
$W^+A(\mu^+\mu^-)$ decay channel from CMS~\cite{1905.07453}
assuming $M_{H^+}-M_A=$ 85~GeV.
Based on the 95\% CLs limits on the signal strength reported in those
analyses, we identify excluded regions of the parameter space as shown in Fig.~\ref{fig:para},
for the two classes of type-I models considered.
The constraints are strongest for models with scenario $\mathcal{A}$ from searches
of $\tau^+\nu_{\tau}$ channel.
The parameter space are less constrained in models with scenario $\mathcal{B}$
by the CMS search of $W^+\mu^+\mu^-$. 
The results agree with those shown in Ref.~\cite{1906.02520} except for
constraints imposed by the ATLAS search of $\tau^+\nu_{\tau}$ that is
not considered therein.
There also exists a recent analysis from CMS on the $W^+\mu^+\mu^-$ channel
using a larger data sample of 139~$fb{^{-1}}$~\cite{ATLAS:2021xhq}.
They report improved limits on 
the signal strength by up to a factor of two comparing with those in~\cite{1905.07453}.
For the decay channel of $W^+b\bar b$ considered in this work,
we plot contours of $B^{sig.}_{H^+}$ in the plane of $\tan\beta$ and $M_{H^+}$ in Fig.~\ref{fig:para}
for the two classes of type-I models.
In models with scenario $\mathcal {A}$ the signal strength can reach at most
0.3\% for the allowed parameter space by previous direct searches, with
a mass of the charged Higgs boson between 100 and 160~GeV.
The new constraints are represented by the solid line which are comparable
to the CMS ones for a mass greater than 150~GeV but weaker
than the ATLAS constraints.
For the class of models with scenario $\mathcal {B}$, with constraints
from previous searches the signal strength can still be larger than
1\%.
From last section the limit on the signal strength is 0.28(0.64)\% at $M_{H^+}$
of 140(110)~GeV.
That excludes the parameter space of $\tan\beta <5\sim 8$ for a mass
up to 145~GeV, representing the strongest constraints from direct searches.
Lastly, we mention that various theoretical and indirect constraints
are generally weaker than those from LHC direct searches for type-I models
as discussed in Refs.~\cite{1706.05980,2103.07484,2106.13656,2201.06890}.
For instance, measurements on electroweak precision observables
indicate either of the mass splits between the charged Higgs boson and the
two additional neutral Higgs bosons should be small~\cite{2103.07484}.
It is possible that the CP-even Higgs boson is lighter than the charged
Higgs boson and contributes via cascade decays similar as the CP-odd Higgs boson.
A scan on full parameter space of type-I 2HDMs combining all relevant decay
channels is desirable which we leave for future investigation.
\noindent \textbf{Summary.} 
We have studied signature of a light
charged Higgs boson produced from top quark pairs at the LHC via its
three-body decays into a $W$ boson and a pair of bottom quarks.
We obtain the first direct limit on signal strength of the decay channel
using the ATLAS measurement at 13 TeV on the relevant final states.
The 95\% CLs limits range from 0.28\% to 0.74\% on $Br(t\rightarrow H^+b)\times Br(H^+\rightarrow W^+b\bar b)$
for a charged Higgs boson with mass between 110 and 160~GeV.
The limits are translated into constraints on the parameter space of 2HDMs of type-I.
We find the strongest constraints on a class of type-I models with a light CP-odd Higgs boson,
compared with  previous direct searches via other decay channels at the LHC. 
We encourage dedicated searches by experimental collaborations for
further improvements.
%

\quad \\
\noindent \textbf{Acknowledgments.} 
The work of JG is supported by the National Natural Science Foundation
of China under Grants No.~11875189 and No.11835005. 
The authors would like to thank Hong-Jian He for useful discussions.
We thank the sponsorship from Yangyang Development Fund.
%
%
\bibliographystyle{apsrev}
\bibliography{ch2hdm}

\end{document}